\documentclass[showpacs,nofootinbib,aps,superscriptaddress,prd,notitlepage,showkeys,10pt]{revtex4}
\usepackage{multirow,dcolumn,bm,bbm,morefloats,color,amsmath,mathrsfs,slashed,multirow,amssymb,bbding,appendix,graphicx}
\usepackage[colorlinks=true, linkcolor=blue, citecolor=blue, urlcolor=blue]{hyperref}
\usepackage{float,subfloat,indentfirst,xcolor,subfigure,gensymb}
\usepackage[sort&compress]{natbib}
\usepackage{txfonts}
\DeclareMathAlphabet{\mathcal}{OMS}{cmsy}{m}{n}
\DeclareSymbolFont{largesymbols}{OMX}{cmex}{m}{n}

\makeatletter
\newcommand*{\rom}[1]{\expandafter\@slowromancap\romannumeral #1@}
\usepackage{ulem}

\allowdisplaybreaks
\makeatother

\begin{document}

\title{Nature of $K^*(1680)$ and $q\bar{q}$-hybrid mixing as the SU(3) partner of $\eta_{1}(1855)$ in the strange sector}
\author{Samee Ullah}
\affiliation{Institute of High Energy Physics, Chinese Academy of Sciences, Beijing 100049, China}
\affiliation{University of Chinese Academy of Sciences. Beijing 100049. China}

\author{Ye Cao}
\affiliation{Southern Center for Nuclear-Science Theory (SCNT), Institute of Modern Physics, Chinese Academy of Sciences, Huizhou 516000, China}

\author{Ming-Xiao Duan}
\affiliation{School of Physics and Electronic Science, Guizhou Normal University, Guiyang 550025, P.R.China}

\author{Hai-Bing Fu}
\affiliation{Institute of High Energy Physics, Chinese Academy of Sciences, Beijing 100049, China}
\affiliation{Department of Physics, Guizhou Minzu University, Guiyang 550025, P.R.China}

\author{Qiang Zhao}
\email{zhaoq@ihep.ac.cn}
\affiliation{Institute of High Energy Physics, Chinese Academy of Sciences, Beijing 100049, China}
\affiliation{University of Chinese Academy of Sciences. Beijing 100049. China}
\affiliation{Center for High Energy Physics, Henan Academy of Sciences, Zhengzhou 450046, China}

\renewcommand{\thefootnote}{\fnsymbol{footnote}}
\footnotetext{Ye Cao, Ming-Xiao Duan, Hai-Bing Fu, and Samee Ullah contribute equally to this work.}

\begin{abstract}
We presents an investigation of the $K^*(1680)$ state in its strong decays into two-body finial states within the flux-tube model and quark pair creation model. Since the charge conjugation parity is not conserved in the strange sector, the conventional $q\bar{q}$ states of $J^{P(C)}=1^{-(-)}$ can mix with the lowest hybrid states with $J^{P(C)}=1^{-(+)}$. Our analysis of the $K^*(1680)$ two-body strong decays indicates that the decay pattern of $K^*(1680)$ cannot be explained by the conventional $q\bar{q}$ scenario. Meanwhile, strong evidence shows  the $q\bar{q}$-hybrid mixing mechanism in the strange sector. The phenomenological consequences of such a mixing are also discussed. Our study can provide a guidance for the future search for hybrid multiplets in experiment at BESIII, LHCb, and Belle-II.

\end{abstract}

\maketitle

\section{Introduction}
The conventional quark model provides a useful framework in classifying hadrons made of quark and/or antiquarks, where mesons are described as bound states of quark and antiquark ($q\bar{q}$), and baryons as three quark systems ($qqq$). This model, though conceptually simple, has yielded remarkable success in explaining the properties and spectrum of hadrons, leveraging the constituent quark dynamics to capture essential features of the hadron structure. Quantum Chromo-Dynamics (QCD), the foundational theory of strong interactions, also predicts the existence of the ``so-called" exotic hadrons with more complicated constituent structures beyond the conventional quark model. These exotic states can provide rich information about the non-perturbative aspects of QCD, and serve as valuable probes for understanding its complex phenomena.

Among the various exotic candidates, hadrons with quantum numbers that defy the conventional quark model are considered strong evidence for the existence of exotic hadrons. Specifically, a hybrid state is composed of constituent quark and antiquark plus a constituent gluon which can naturally exhibit quantum numbers beyond the conventional $q\bar{q}$ scenario. For instance, for the lowest hybrid systems with the orbital angular momentum 0, the quantum numbers $J^{P(C)} = 0^{-(+)}/1^{-(\pm)}/2^{-(+)}$ can be accessed, where $J^{PC}=1^{-+}$ cannot be accessed by the conventional $q\bar{q}$ configuration. This particular class of hadrons has sparked significant attention from both experiment and theory, driving continued explorations of the hadron structures and underlying dynamics. In 2009 the COMPASS Collaboration reported the $J^{PC}=1^{-+}$ candidate with isospin 1, i.e. $\pi_1(1600)$, through the partial wave analysis~\cite{COMPASS:2009xrl}. Their analysis also clarified the evidence for $\pi_1(1400)$ which seems to be caused by interference from $\pi_1(1600)$. In 2022, the BESIII Collaboration reported the first observation of the $1^{-+}$ isoscalar hybrid candidate $\eta_1(1855)$ in a partial wave analysis of $J/\psi \to \gamma \eta_1(1855) \to \gamma \eta \eta'$~\cite{BESIII:2022riz, BESIII:2022iwi}. The mass and total decay width of this state are $(1855\pm9^{+6}_{-1})~{\rm MeV}$ and $(188\pm18^{+3}_{-8})~{\rm MeV}$, respectively. This breakthrough opens up a valuable opportunity to gain a better and deeper insight of these mysterious hadron species, and has initiated a lot of attention from theory~\cite{Qiu:2022ktc,Chen:2022qpd,Wan:2022xkx,Shastry:2022mhk,Dong:2022cuw,Wang:2022sib,Chen:2022isv,Yan:2023vbh,Shastry:2023ths,Swanson:2023zlm,Chen:2023ukh,Dong:2022otb,Tan:2024grd}. In fact, an immediate question on the hybrid states is about their partners. For instance, in the light quark sector one would expect the existence of a flavor nonet $q\bar{q}\tilde{g}$. An immediate investigation of the hybrid nonet after the observation of $\eta_1(1855)$ can be found in Ref.~\cite{Qiu:2022ktc}. Historically, experimental efforts on the search of the hybrid states can be found in the literature~\cite{SLACHybridFacilityPhoton:1991oug, VES:1993scg, Lee:1994bh, Aoyagi:1993kn, E852:1997gvf, E852:1999xev}. Comprehensive reviews of the relevant issues can be found in Refs.~\cite{Klempt:2007cp, Meyer:2015eta}.

The possible existence of the hybrid nonet has a strong support from the lattice QCD simulations~\cite{Dudek:2013yja, Dudek:2011bn}. With $\pi_1(1600)$ assigned as the isovector and $\eta_1(1855)$ as one of the isoscalar, there are still two pieces of information to be filled into the jigsaw puzzle. Namely, the isospin $1/2$ partner with strangeness $\pm 1$ and the other isoscalar state $\eta_1'$. In Ref.~\cite{Qiu:2022ktc} possible solutions for the nonet spectrum based on the Gell-Manm-Okubo mass relation were discussed. An interesting feature with the isospin $1/2$ partners is that they do not have a fixed $C$-parity. With $J^P=1^-$, such a hybrid state has the same quantum numbers as the $q\bar{q}$ state. Thus, it may mix with the $1^-$ states in the same mass regime.

The strange $I=1/2$ partner can be assigned to $K^*(1680)$ which is the only strange vector meson found in the vicinity of the mass region $1.6-1.9$ GeV~\cite{ParticleDataGroup:2020ssz}. Given that $\pi_1(1600)$ is the isovector state of the lowest $1^{-+}$ hybrid, $K^*(1410)$ is too light to be its strange partner. Actually, $K^*(1410)$ appears also to be too light for a $2^3S_1$ state taking into account its isovector partner $\rho(1450)$~\cite{ParticleDataGroup:2020ssz}. In Ref.~\cite{Barnes:2002mu}, it was suggested that the low mass of $K^*(1410)$ might be due to the presence of additional hybrid mixing states.

It is possible to assign $K^*(1680)$ as the $1^3D_1$ vector in the strange sector. However,  the relative branching fraction of $K^*(1680)^-$ to $K^-\eta$ and $K^-\pi^0$ measured by the Belle Collaboration, $0.11\pm0.02({\rm stat})^{+0.06}_{-0.04}({\rm syst})\pm0.54({\cal B}_{\rm PDG})$~\cite{Belle:2020fbd}, seems not to be consistent with the theoretical predictions $(\approx 1)$ under the assumption that $K^*(1680)$ is a pure $1^3D_1$ state~\cite{Burakovsky:1997,Pang:2017dlw}.  In Ref.~\cite{Burakovsky:1997} the relationship between the $K^*(1410)$ and $K^*(1680)$ mesons, and the $D$-wave spectrum have been investigated and it also shows that $K^*(1680)$ cannot be a pure $^3D_1$ state.

The above issues leave a possibility that $K^*(1680)$ could be a candidate for the strange hybrid or it may be a mixing state between the $q\bar{q}$ and hybrid configuration~\cite{Qiu:2022ktc}.

In this work we explore the possibility for $K^*(1680)$ to be the strange hybrid of the $1^{-(+)}$ nonet. Meanwhile, it should be realized that the hybrid $K^*$ can mix with the conventional vector state. To clarify this, we also consider the impact of the $q\bar{q}$ mixing with the hybrid on the decay of $K^*(1680)$. The idea is that given $K^*(1680)$ to be a conventional $q\bar{q}$ state in the quark model, its two-body decays into different channels can be reasonably described by the quark pair creation model (QPC)~\cite{Micu:1968mk,LeYaouanc:1973ldf, LeYaouanc:1972vsx, LeYaouanc:1974cvx, LeYaouanc:1977gm,Roberts:1992esl}, which is also known as the $^3P_0$ ($n^{2S+1}L_J$) model. Since the QPC model has been broadly applied to various decay processes and turned out to be successful~\cite{Lu:2006ry, Yu:2011ta, Ye:2012gu,  Lu:2016bbk,Chen:2007xf, Chen:2016iyi, Zhao:2016qmh}, significant deviations from the QPC model expectation may indicate non-trivial mechanisms or structures different from the $q\bar{q}$ scenario. It is a reasonable assumption that as the strange hybrid of the $J^{P(C)}=1^{-(+)}$ nonet the two-body hadronic decays of $K^*(1680)$ should be different from the $q\bar{q}$ decays. We then try to explore whether such a possibility does exist.

While there have been various phenomenological studies of the hybrid states in the literature~\cite{Horn:1977rq,Barnes:1982zs,Page:1998gz, Close:2003af}, the flux-tube model (FT) ~\cite{Isgur:1984bm} provides a dynamic connection for the decay mechanism of a hybrid state with a $q\bar{q}$ meson in the QPC model. For the decay of a hybrid in the FT model the collinear mode describing the motion of the constituent gluon along the direction between $q$ and $\bar{q}$ can be regarded as being equivalent to the QPC model if the multi-soft-gluon exchanges are included in the decay. In contrast, the transverse mode of the gluon motion should manifest itself as a different mechanism from the QPC model since contributions from the gluon degrees of freedom can contribute to the quantum numbers in addition to the $q\bar{q}$ sub-system. In this sense, we can parametrize out the transitions of both the $q\bar{q}$ and hybrid configurations and examine whether the present experimental data indicate some evidence for the hybrid configuration.

The remaining parts of the paper are organized as follows: In Sec.~\ref{Sec:II} we will present our theoretical framework. In Sec.~\ref{Sec:III} numerical results will be given. Comparisons with the available experimental data and discussions about the underlying mechanisms will be presented. A brief summary will be given in Sec.~\ref{Sec:IV}.

\section{Theoretical Framework}\label{Sec:II}

Our strategy of studying the property of $K^*(1680)$ is based on the following considerations: (i) As mentioned earlier that the two-body decays of an initial $q\bar{q}$ can be reasonably described in the QPC model. It thus allows us to investigate the $q\bar{q}$ scenario for $K^*(1680)\to PP$ and $VP$ as a conventional $q\bar{q}$ state and establish the relation among these decay channels via the SU(3) flavor symmetry. By assigning $K^*(1680)$ to be the first radial excitation state of $2^3S_1$ or orbital excitation state of $1^3D_1$, we will examine whether it can be described by the $q\bar{q}$ scenario. (ii) By introducing the hybrid scenario we will also investigate the decay pattern of the initial hybrid $K^*(1680)\to PP$ and $VP$ with the collinear and transverse mode considered.  
(iii) We then consider that $K^*(1680)$ as the mixing state of the conventional $q\bar{q}$ state and hybrid. The mixing angle will be fitted by the overall description of all the two-body decay channels of $K^*(1680)\to PP$ and $VP$. 

At the hadronic level the decay of an initial vector meson into $PP$ and $VP$ can be described by effective Lagrangians. Given $V$ and $P$ standing for the ground state vector and pseudoscalar mesons, the SU(3) flavor symmetry may allow us to set up connections among all the $PP$ and $VP$ decay channels.  
The following effective Lagrangians are adopted:
\begin{align}
    \mathcal{L}_{VPP}&= i g_{VPP} {\rm Tr} [(P\partial_{\mu}P- \partial_{\mu}P P)V^{\mu}],
    \\
    \mathcal{L}_{VVP}&= \frac{1}{m_V} g_{VVP}\epsilon_{\alpha\beta\mu\nu} {\rm Tr} [\partial^\alpha V^\mu \partial^\beta V^\nu P],
\end{align}
where $g_{VPP}$ and $g_{VVP}$ are the dimensionless coupling strengths. Again, $V$ and $P$ stand for the vector and pseudoscalar field for the ground state SU(3) nonet and have the following form
\begin{widetext}
\begin{align}
P & = \left(
\begin{array}{*{20}{c}}
{\dfrac{{\sin {\alpha _P}\eta' + \cos {\alpha _P}\eta  + {\pi ^0}}}{{\sqrt 2 }}}&{{\pi ^ + }}&{{K^ + }}\\
\pi^- &{\dfrac{{\sin {\alpha _P}\eta' + \cos {\alpha _P}\eta  - {\pi ^0}}}{{\sqrt 2 }}}&{{K^0}}\\
{{K^ - }}&{{{\bar K}^0}}&{\cos {\alpha_P}\eta' - \sin {\alpha _P}\eta }
\end{array}
\right),
\end{align}
and
\end{widetext}
\begin{align}
V &= \left( \begin{array}{*{20}{c}}
\dfrac{\omega  + \rho ^0}{\sqrt 2}&\rho^+&K^{*+}\\
\rho^-&\dfrac{\omega - \rho ^0}{\sqrt 2}&K^{*0}\\
{{K^{* - }}}&{{{\bar K}^{*0}}}&\phi
\end{array} \right).
\end{align}
In the vector sector the ideal mixing between $\omega=(u\bar u + d\bar d)/\sqrt 2$ and $\phi = s\bar s$ is adopted. As for the $\eta$ and $\eta'$, they can be expressed on the quark-flavor basis as $\eta = \cos \alpha_P n\bar n - \sin\alpha_P s\bar s$ and $\eta' = \sin \alpha_P n\bar n + \cos\alpha_P s\bar s$, where $n\bar n \equiv (u\bar u + d\bar d)/\sqrt2$, and the mixing angle $\alpha_P \equiv \arctan\sqrt 2 + \theta_p$ with $\theta_p$ the flavor singlet and octet mixing angle. For the two-body hadronic decays, most of the decays of $K^*(1680) \to PP$ and $VP$ are allowed by the kinematics. We will focus on the $PP$ and $VP$ decay channels taking into account the experimental situation. The hadronic level transition amplitudes in effective Lagrangian approach (ELA) are as follows:
\begin{align}
i\mathcal{M}_{VPP}^{\text{ELA}}&= ig_{VPP}(p_B-p_C)_{\mu}\epsilon_A^{\mu}
\nonumber\\
&= i g_{VPP} 2 |\boldsymbol p_B|
\label{Eq:HadronicAmplitudesELAVPP}
\\
i\mathcal{M}_{VVP}^{\text{ELA}}&= \frac{1}{m_A}ig_{VVP}\epsilon_{\alpha\beta\mu\nu} p_A^{\alpha}p_B^{\beta}\epsilon_A^{\mu}{\epsilon_B^{\nu}}^*
\nonumber\\
&=i g_{VVP}m_A |\boldsymbol p_B|.
\label{Eq:HadronicAmplitudesELAVVP}
\end{align}

\begin{figure}[t]
\centering
\includegraphics[width=0.32\textwidth]{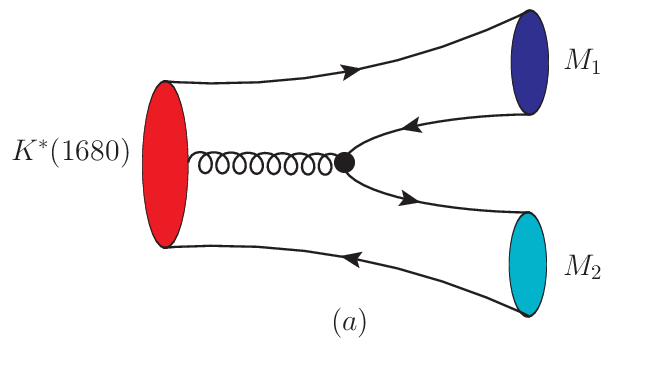}
\includegraphics[width=0.32\textwidth]{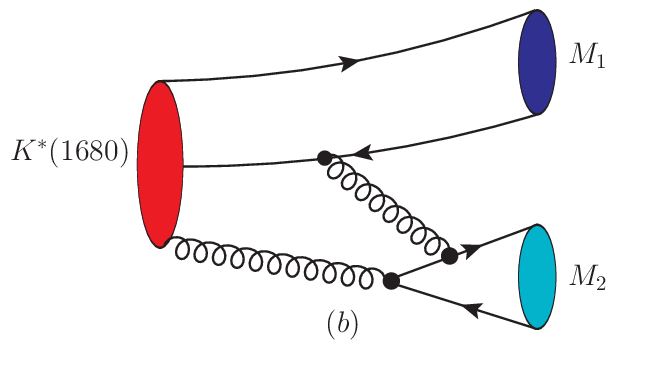}
\caption{Schematic illustrations for the two-body hadronic decays of hybrid states. (a) is dominated by the collinear mode, and (b) is dominated by the transverse mode. }
\label{Fig:FT}
\end{figure}
The above defined coupling constants can have contributions from the decay of the conventional $q\bar{q}$ component via the QPC model, and from the decay of the hybrid component via the FT model.

As discussed earlier, the FT model actually contains the QPC model. Its collinear mode shares similar dynamics as the QPC model in the decay. Since we only consider the decay of a fixed process, i.e. $K^*(1680)\to VP$ and $PP$, we assume that the collinear mode can be described by the QPC model. Meanwhile, the transverse mode will be driven by a different  mechanism. These are two independent mechanisms for the hybrid decays, and can be illustrated by Fig.~\ref{Fig:FT} (a) and (b), respectively. At the quark level  Fig.~\ref{Fig:FT} (a) dominates the collinear mode and describes the breaking of the flux-tube string by creating a pair of $q\bar{q}$ to combine with the initial quark and antiquark. With multi-gluon exchanges in the course of the hadronization Fig.~\ref{Fig:FT} (a) will behave similarly as the QPC transition.  It allows us to evaluate the coupling of the collinear mode decay by the QPC model and unify these two mechanisms in the same framework.

The hybrid decay via the transverse mode will dominate the process of Fig.~\ref{Fig:FT} (b).
The created quark pair from the transverse mode will recoil the initial quark anti-quark via the transverse flux motion. Then, with a soft gluon exchange to neutralize the color of the recoiled $q\bar{q}$ pair, the process of Fig.~\ref{Fig:FT} (b) will play a crucial and unique role in the decay channels involving the production of flavor-singlet hadrons. For conventional $q\bar{q}$ decays, a similar process as Fig.~\ref{Fig:FT} (b) may also occur. However, it is usually suppressed by the Okubo-Zweig-Iizuka (OZI) rule. In contrast, in the decay of hybrid Fig.~\ref{Fig:FT} (b) is not necessarily suppressed~\cite{Kokoski:1985is}.

With the above consideration we can define the transition amplitude for a $1^{-+}$ hybrid of $q\bar{q}\tilde{g}$ decaying into two conventional mesons $PP$ and $VP$ as follows~\cite{Qiu:2022ktc}
\begin{align}
\begin{split}
    &\mathcal{M}_a^{VPP}= \langle(q_1\bar{q_4})_{P} (q_3\bar{q_2})_{P}|\hat V_C|q_1\bar{q_2}\tilde{g}\rangle \equiv g_{1}^{VPP}|\boldsymbol p_B |,
    \\
    &\mathcal{M}_b^{VPP}= \langle(q_1\bar{q_2})_{P} (q_3\bar{q_4})_{P}|\hat V_T|q_1\bar{q_2}\tilde{g}\rangle \equiv g_{2}^{VPP}|\boldsymbol p_B |, \\
    &\mathcal{M}_a^{VVP}= \langle(q_1\bar{q_4})_{P(V)} (q_3\bar{q_2})_{V(P)}|\hat V_C|q_1\bar{q_2}\tilde{g}\rangle \equiv g_{1}^{VVP}|\boldsymbol p_B |,
    \\
    &\mathcal{M}_b^{VVP}= \langle(q_1\bar{q_2})_{P(V)} (q_3\bar{q_4})_{V(P)}|\hat V_T|q_1\bar{q_2}\tilde{g}\rangle \equiv g_{2}^{VVP}|\boldsymbol p_B |, 
    \end{split}
\end{align}
Where $\mathcal{M}_a$, $\mathcal{M}_b$ are amplitudes corresponding to the collinear and transverse flux excitations, and $\hat{V}_C$ and $\hat{V}_T$ are the corresponding interaction potentials. $|\boldsymbol p_B |$ is three-momentum of the final-state  meson in the c.m. frame of the initial hybrid.

By matching the transition amplitudes defined at the quark level to those defined at the hadronic level, we can have the following relations for $K^*(1680)\to PP$ and $VP$ as the hybrid state:
\begin{align}
    \begin{split}
        \mathcal{M}^{\text{FT}}_{PP}&=2g_{PP}^{\text{FT}}|\boldsymbol{p}_B|,\ \ \ \ PP=K^+\pi^0,K^+\eta,K^+\eta^{\prime},\\
        \mathcal{M}^{\text{FT}}_{VP}&=g_{VP}^{\text{FT}}|\boldsymbol{p}_B|,\ \ \ \ \ VP=K^{*+}\pi^0,\rho^0K^+,\phi K^+,\omega K^+,K^{*+}\eta \ .
    \end{split}
\end{align}
The above relations will allow a unified description of the decay processes with the two coupling constants $g_1$ and $g_2$, with which the hadronic couplings $g_{PP/VP}^{\text{FT}}$ for the 8 decay channels of $K^{*+}(1680)\to PP$ and $VP$ can be parameterized as
\begin{align}
 g_{PP/VP}^{\text{FT}}=g_{PP/VP}^{\text{FT(C)}}(g_1)+g_{PP/VP}^{\text{FT(T)}}(g_2) \ .
\end{align}
The specific expression for each channel can be obtained:
\begin{align}
        g_{K^+\pi^0}^{\text{FT(C)}}&=\frac{1}{\sqrt{2}}g_1^{K^+\pi^0},&g_{K^+\pi^0}^{\text{FT(T)}}&=0, \label{eq:FT-coupling i}\\
        g_{K^+\eta}^{\text{FT(C)}}&=\bigg(\frac{1}{\sqrt{2}}\text{cos}\alpha_P-R\text{sin}\alpha_P\bigg)g_{1}^{K^+\eta},&g_{K^+\eta}^{\text{FT(T)}}&=\bigg(\sqrt{2}\text{cos}\alpha_P-R\text{sin}\alpha_P\bigg)\delta g_{1}^{K^+\eta},\\
        g_{K^+\eta^{\prime}}^{\text{FT(C)}}&=\bigg(\frac{1}{\sqrt{2}}\text{sin}\alpha_P+R\text{cos}\alpha_P\bigg)g_{1}^{K^+\eta^{\prime}},&g_{K^+\eta^{\prime}}^{\text{FT(T)}}&=\bigg(\sqrt{2}\text{sin}\alpha_P+R\text{cos}\alpha_P\bigg)\delta g_{1}^{K^+\eta^{\prime}},\\
        g_{K^{*+}\pi^0}^{\text{FT(C)}}&=\frac{1}{\sqrt{2}}g_1^{K^{*+}\pi^0},&g_{K^{*+}\pi^0}^{\text{FT(T)}}&=0\\
        g_{K^+\rho^0}^{\text{FT(C)}}&=\frac{1}{\sqrt{2}}g_1^{K^+\rho^0},&g_{K^+\rho^0}^{\text{FT(T)}}&=0,\\
        g_{K^+\omega}^{\text{FT(C)}}&=\frac{1}{\sqrt{2}}g_1^{K^+\omega},&g_{K^+\omega}^{\text{FT(T)}}&=\frac{2}{\sqrt{2}}\delta g_1^{K^+\omega},\\
        g_{K^+\phi}^{\text{FT(C)}}&=Rg_1^{K^+\phi},&g_{K^+\phi}^{\text{FT(T)}}&=R\delta g_1^{K^+\phi},\\
        g_{K^{*+}\eta}^{\text{FT(C)}}&=\bigg(\frac{1}{\sqrt{2}}\text{cos}\alpha_P-R\text{sin}\alpha_P\bigg)g_{1}^{K^{*+}\eta},&g_{K^{*+}\eta}^{\text{FT(T)}}&=\bigg(\sqrt{2}\text{cos}\alpha_P-R\text{sin}\alpha_P\bigg)\delta g_{1}^{K^{*+}\eta},
        \label{eq:FT-coupling f}
\end{align}
where we label the decay channels in the superscript of $g_1$ and $g_2$ to highlight the possible differences arising from their distinct spatial convolutions in different decay channels. The ratio $\delta \equiv g_2/g_1$ stand for the relevant strength of the two couplings $g_{1,2}$. For the decay channels where the final states do not have isoscalars, $g_2$ should vanish. For instance, in $K^*(1680)\to K\pi$ only Fig.~\ref{Fig:FT} (a) will contribute. In this sense, it is difficult to distinguish between the $q\bar{q}$ and $q\bar{q}\tilde{g}$ scenario.

The assumption that the collinear mode can be described by the QPC model makes it possible to quantify this amplitude by explicit calculations. 
To proceed, we employ the QPC to extract the transition amplitudes for the $q\bar{q}$ scenario. These amplitudes can provide an estimate of the collinear-mode amplitudes for the hybrid scenario. Then, with the introduction of the relative coupling strength $\delta$, one can evaluate all the decay channels of $K^*(1680)\to PP$ and $VP$.

The QPC amplitude for a $q\bar{q}$ decay can be expressed as the following form:
\begin{align}
    \begin{split}
        \mathcal{M}_{PP/VP}^{\text{QPC}}=\gamma C^{\text{QPC}}_{PP/VP}I^{\text{QPC}}_{PP/VP},
        \label{eq:M^QPC}
    \end{split}
\end{align}
where $\gamma$ is the coupling strength creating the $q\bar{q}$ pair from vacuum and $C_{PP/VP}^{\text{QPC}}$ is a constant coming from all the Clebsch-Gordan coefficients, spin and flavor overlaps and energy term $\sqrt{E_A E_B E_C}$;  $I_{PP/VP}^{\text{QPC}}$ is the spatial integral from the initial and final-state wave function convolution, and is the function of the outgoing momentum $\boldsymbol p_B$. In comparison with the hadronic level transition amplitudes in the ELA for Eqs.~\eqref{Eq:HadronicAmplitudesELAVPP} and \eqref{Eq:HadronicAmplitudesELAVVP}, the QPC amplitude can also be expressed as
\begin{align}
    \begin{split}
        \mathcal{M}_{PP}^{\text{QPC}}&=2g_{PP}^{\text{QPC}}|\boldsymbol{p}_B|,\ \ \ \ PP=K^+\pi^0,K^+\eta,K^+\eta^{\prime}\\ 
        \mathcal{M}_{VP}^{\text{QPC}}&=g_{VP}^{\text{QPC}}|\boldsymbol{p}_B|,\ \ \ \ \ VP=K^{*+}\pi^0,\rho^0K^+,\phi K^+,\omega K^+,K^{*+}\eta.
        \label{eq:M^QPC ELA}
    \end{split}
\end{align}
By matching Eq. (\ref{eq:M^QPC}) with Eq. (\ref{eq:M^QPC ELA}), the effective coupling constant $g_{PP/VP}^{\text{QPC}}$ for each decay channel can be determined from the QPC amplitude,
\begin{align}
    \begin{split}
        g^{\text{QPC}}_{PP}&=\frac{\gamma}{2}C_{PP}^{\text{QPC}}\tilde{I}_{PP}^{\text{QPC}}(|\boldsymbol{p}_B|),\\
        g^{\text{QPC}}_{VP}&=\gamma C_{VP}^{\text{QPC}}\tilde{I}_{VP}^{\text{QPC}}(|\boldsymbol{p}_B|),
        \label{eq:matching coupling}
    \end{split}
\end{align}
where the superscript ``QPC" with $g_{PP}$ and $g_{VP}$ indicates that the effective couplings are extracted from the QPC model.

Consider the mixing of the conventional $q\bar{q}$ and hybrid component within $K^*(1680)$ with the mixing angle $\zeta$, 
\begin{equation}
    |K^*(1680)\rangle=\cos\zeta |q\bar{q}(n^{2S+1}L_{J=1})\rangle+\sin\zeta |q\bar{q}\tilde{g}\rangle \ ,
\end{equation}
where $\zeta$ is the mixing angle; $n$, $L$, $S$, and $J$ are the radial quantum number, total spin, relative orbital angular momentum between quark and anti-quark, and the total angular momentum, respectively, for the $q\bar{q}$ system~\footnote{We assume that the $q\bar{q}$-hybrid mixing can be accommodated by a unitary transformation. Thus, such a mixing pattern has actually predicted a partner with a wavefunction of $|K^{*\prime}\rangle=-\sin\zeta |q\bar{q}(n^{2S+1}L_{J=1})\rangle+\cos\zeta |q\bar{q}\tilde{g}\rangle$. Due to lack of experimental information, we leave the relevant discussion to be addressed later.}. The total amplitudes can then be expressed as
\begin{align}\label{tot-trans-amp}
    \begin{split}
\mathcal{M}_{PP/VP}^{\text{Tot}}&=\text{cos}\zeta \mathcal{M}_{PP/VP}^{\text{QPC}}+\text{sin}\zeta \mathcal{M}_{PP/VP}^{\text{FT}},
    \end{split}
\end{align}
with $\zeta=0/\pi$, the amplitudes describe the initial state $K^*(1680)$ as the conventional $q\bar{q}$ meson; With $\zeta=\pi/2$, the amplitudes describe $K^*(1680)$ as the pure hybrid state. The total amplitude for each decay channel is given as follows:
\begin{align}
    \begin{split}
        \mathcal{M}_{PP}^{\text{tot}}&=2|\boldsymbol{p}_B|\Bigg[g^{\text{QPC}}_{PP}\text{cos}\zeta+\bigg(g_{PP}^{\text{FT(C)}}+g_{PP}^{\text{FT(T)}}\bigg)\text{sin}\zeta\Bigg],PP=K^+\pi^0,K^+\eta,K^+\eta^{\prime},\\
        \mathcal{M}_{VP}^{\text{tot}}&=|\boldsymbol{p}_B|\Bigg[g^{\text{QPC}}_{VP}\text{cos}\zeta+\bigg(g_{VP}^{\text{FT(C)}}+g_{VP}^{\text{FT(T)}}\bigg)\text{sin}\zeta\Bigg],VP=K^{*+}\pi^0,\rho^0K^+,\phi K^+,\omega K^+,K^{*+}\eta.
        \label{eq:M^tot}
    \end{split}
\end{align}
As discussed before, the coupling strength for the collinear mode can be approximated by the QPC coupling strength, i.e., $g_{PP/VP}^{\text{FT(C)}}=g_{PP/VP}^{\text{QPC}}$. Thus, $g_1^{PP/VP}$ for each decay channel can be extracted from the QPC model. At this point, $g_1^{PP/VP}$ is a function of $\gamma$~\footnote{It should be addressed that although the collinear mode cannot be distinguished from the QPC model, it is likely that the collinear mode hybrid decay coupling $\gamma$ defined in analogue with the QPC model coupling, should take a smaller value than that adopted in the QPC model. We will discuss this later in Sec. III.}, the constituent quark masses $(m_s, m_q)$, and the harmonic oscillator parameters $(\beta_{P}, \beta_{V}, \beta_{K^*(1680)})$ from the wave function. After considering $\alpha_P=42^\circ$, in principle, we are left with only two free parameters, i.e. $\delta$ and $\zeta$ in the transition amplitude. They can be constrained by the experimental data, and their values can provide physical information regarding the composition of $K^*(1680)$.

\begin{figure}[t]
\centering
\includegraphics[width=0.32\textwidth]{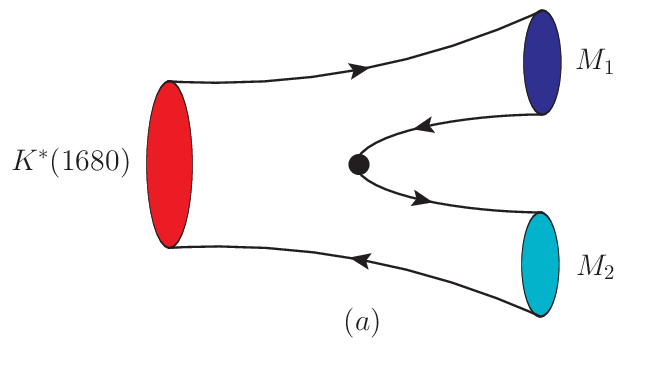}
\includegraphics[width=0.32\textwidth]{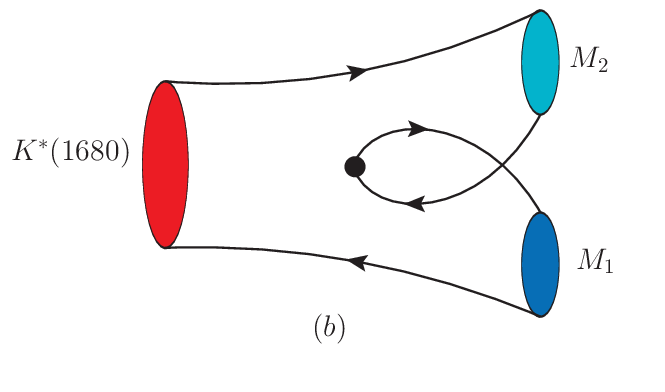}
\caption{Illustrations of the strong decays of $K^*(1680)$ as a $q\bar{q}$  into $PP$ or $PV$ in the QPC model. The collinear mode of a hybrid decay via Fig.~\ref{Fig:FT} (a) cannot be distinguished from the QPC mechanism shown here. }
\label{Fig:QPC}
\end{figure}

To keep the integrity of the formulation, we briefly summarize the QPC model here. In the QPC model, the quark pair creation process from vacuum can be described as
\begin{align}
    \begin{split}
        \hat{T}=-3\gamma\sum_m\langle 1m;1-m|00\rangle{{(2\pi)^{3/2}}}\int d^3\boldsymbol{p}_3d^3\boldsymbol{p}_4\delta^3(\boldsymbol{p}_3+\boldsymbol{p}_4)y_{1m}(\frac{\boldsymbol{p}_3-\boldsymbol{p}_4}{2})\chi_{1-m}^34\phi_0^{34}\omega_0^{34}b_{3i}^{\dagger}(\boldsymbol{p}_3)d_{4j}^{\dagger}(\boldsymbol{p}_4).
    \end{split}
\end{align}
Here the momenta of created quarks and anti-quark are integrated over all values, with a condition that their total momentum sum is zero; Subscripts $i$, $j$ are the color SU(3) indices of the created quark and antiquark; ${\boldsymbol p}_i$ denotes the $i$-th quark (antiquark) three-vector momentum; $\phi_0^{34}=(u\bar{u}+d\bar{d}+s\bar{s})/\sqrt{3}$, $\omega_0^{34}=\frac{1}{\sqrt{3}}\delta_{ij}$ and $\chi_{1-m}^{34}$ are the flavor singlet, color singlet and spin triplet wave functions of the created $q_3\bar{q}_4$ pair, respectively; the first solid harmonic polynomial ${\mathcal Y}_{1m}(\boldsymbol p)\equiv|\boldsymbol p|^1Y_{1m}(\theta,\phi)$ reflects the $P$-wave momentum-space distribution of the $q_3\bar{q}_4$ pair; $\gamma$ is a dimensional constant describing the strength of quark pair creation from vacuum and can be extracted by fitting data. As mentioned earlier, in the collinear mode transition of the hybrid, when the flux tube breaks, the multi-soft-gluon exchanges between the initial and final created light quark-anti-quark pairs will wipe away the information about the initial gluelump. This makes it indistinguishable between the hybrid collinear mode transition and QPC model in the two-body hadronic decays. Taking into account the multi-gluon exchange effects, the coupling $\gamma$ should be smaller than  the values for the conventional QPC model. In the center-of-mass (c.m.) frame of initial meson $K^*(1680)$, the helicity amplitude can be written as
\begin{align}
    \begin{split}
        &\mathcal{M}^{M_{J_A}M_{J_B}M_{J_C}}\\
        &=\gamma\sqrt{8E_AE_BE_C}{(2\pi)^{3/2}}\sum\langle L_AM_{L_A};S_AM_{S_A}|J_AM_{J_A}\rangle\langle L_BM_{L_B};S_BM_{S_B}|J_BM_{J_B}\rangle\\
        &\times\langle L_CM_{L_C};S_CM_{S_C}|J_CM_{J_C}\rangle\langle 1m;1-m|00\rangle\langle\chi_{S_CM_{S_C}}^{32}\chi_{S_BM_{S_B}}^{14}|\chi_{S_AM_{S_A}}^{12}\chi_{1-m}^{34}\rangle\\
        &\times\Bigg[\langle\phi_{C}^{32}\phi_{B}^{14}|\phi_{A}^{12}\phi_{0}^{34}\rangle I_{M_{L_B}M_{L_C}}^{M_{L_A}m}(\boldsymbol{p}_B,m_1,m_2,m_3)\\
        &+(-1)^{1+S_A+S_B+S_C}\langle\phi_{C}^{14}\phi_{B}^{32}|\phi_{A}^{12}\phi_{0}^{34}\rangle I_{M_{L_B}M_{L_C}}^{M_{L_A}m}(-\boldsymbol{p}_B,m_2,m_1,m_3)\Bigg],
    \end{split}
\end{align}
where symbol $\sum$ is the summation over $M_{L_A}$, $M_{S_A}$, $M_{L_B}$, $M_{S_B}$, $M_{L_C}$, $M_{S_C}$ and $m$. Meanwhile, the wave function convolution in the momentum space have the following form,
\begin{align}
    \begin{split}
        I_{M_{L_B}M_{L_C}}^{M_{L_A}m}(\boldsymbol{p}_B,m_1,m_2,m_3)&=\int\boldsymbol{p}_3\psi_B^*(\frac{m_4}{m_1+m_4}\boldsymbol{p}_B+\boldsymbol{p}_3)\psi_C^*(\frac{m_3}{m_2+m_3}\boldsymbol{p}_B+\boldsymbol{p}_3)\psi_A(\boldsymbol{p}_B+\boldsymbol{p}_3)y_{1m}(\boldsymbol{p}_3).
    \end{split}
\end{align}
which serves as the form factor defined at the hadronic level. The wave function in the momentum-space has the following expression,
\begin{align}
\psi_{nlm}(\boldsymbol p, \beta) & =(-1)^n (-i)^l \sqrt{\frac{2n!} {(n + l + 1/2 )!}} \left(\frac{\boldsymbol p}{\beta}\right)^l \frac{1}{\beta^{3/2}}
e^{-\frac{\boldsymbol p^2}{2\beta^2}} L_n^{l+1/2}\left(\frac{\boldsymbol p^2}{\beta^2}\right) Y_{lm}(\theta,\phi)
\nonumber\\
&\equiv R_{nl}(p)Y_{lm}(\theta,\phi),
\end{align}
where the $L_n^a(x)=\sum_{k=0}^{n}(-1)^kC_{n+a}^{n-k}\frac{x^k}{k!}$ denotes the generalized Laguerre polynomial; $Y_{lm}(\theta,\phi)$ is the spherical harmonic function. Note that $\psi_{nlm}$ is the wave function for quark internal motion on the basis of harmonic oscillator (HO), and $\beta$ is the HO strength. The wave function describing the c.m. motion is $\delta^3(\boldsymbol p-\boldsymbol p^{\prime})$. 
The corresponding normalization conditions are
\begin{align}
\begin{split}
&\int _0^{\infty}dp[R_{nl}(p)]^2p^2=1,\\
&\int_0^{\pi}\int_0^{2\pi} d\theta d\phi Y_{l_1m_1}^*(\theta,\phi)Y_{l_2m_2}(\theta,\phi)=\delta_{l_1l_2}\delta_{m_1m_2}.
\end{split}
\end{align}
With the relativistic phase space, the decay width in the c.m. frame ~\cite{Blundell:1995ev}
\begin{equation}
    \Gamma= \frac{|\boldsymbol{p}_B|}{8\pi M_{A}^{2}} \frac{s}{\left(2 J_{A}+1\right)} \sum_{M_{J_{A}}, M_{J_{B}}, M_{J_{C}}}\left|{\cal M}^{M_{J_{A}} M_{J_{B}} M_{J_{C}}}\right|^{2},
\end{equation}
where $|\boldsymbol p_B |=\sqrt{[m_A^2-(m_B+m_C)^2][m_A^2-(m_B-m_C)^2]}/(2m_A)$ is the three momentum of the final-state particle in the c.m. frame; $J_A$ is the spin of the initial state, and $s=1/(1+\delta_{BC})$ is a statistical factor which is needed if $B$ and $C$ are identical particles. In the numerical studies we will investigate the possibilities that $K^*(1680)$ could be assigned as either the radial excitation states $|q\bar{q}(2^3S_1)\rangle$ and $|q\bar{q}(3^3S_1)\rangle$, or the $D$-wave state $|q\bar{q}(1^3D_1)\rangle$.

\section{Results and discussions}\label{Sec:III}

Proceeding to the numerical study, we first investigate the possible assignment of $K^*(1680)$ as the conventional $q\bar{q}$ state in the quark model. For the vector $K^*$ spectrum, as discussed in the Introduction that $K^*(1410)$ could be the candidate for the first radial excitation $|q\bar{q}(2^3S_1)\rangle$ though its mass is too light if one takes into account $\rho(1450)$ as its $I=1$ partner~\cite{ParticleDataGroup:2020ssz}. While one may need to understand why $K^*(1410)$ has such a low mass, alternative possibilities are that $K^*(1680)$ could the second radial excitation state $|q\bar{q}(3^3S_1)\rangle$, or the $D$-wave state $|q\bar{q}(1^3D_1)\rangle$. We will discuss these possibilities with the numerical results later.

Note that we have only five data points from experiment to fit, but have nine parameters in the formalism. One also notices that these experimental measurements of the branching ratios and the total width of $K^*(1680)$ bare quite large uncertainties. Thus, when doing the numerical calculation, it is necessary to fix some of the input parameters in order to reduce the number of parameters. Among these parameters, the constituent quark masses ($m_u=m_d=m_q=300$ MeV, $m_s=500$ MeV), harmonic oscillator strengths ($\beta_P=550$ MeV, $\beta_V=450$ MeV), coupling strength $\gamma_1$ for the QPC model, $\alpha_P=42^\circ$ for the $\eta$ and $\eta'$ mixing, and the SU(3) flavor symmetry breaking factor $R=m_q/m_s$ are fixed in the calculation. By fixing those quark model parameters, it means that we set up some limit for the pattern from the NRCQM. We then examine whether some of these patterns can be accounted for in the NRCQM, and whether the $q\bar{q}$-hybrid mixing can explain the deviations from the NRCQM expectations.

The rest free parameters include the hybrid  couplings $g_1$, $g_2$ and the $q\bar{q}$-hybrid mixing angle $\zeta$, which  will be determined by the numerical fitting. Alternatively, we can define $g_1$ and the relative strength $\delta=g_2/g_1$ as the two independent coupling parameters and $g_1$ will be connected to the QPC coupling via parameter $\gamma_2$ through the relations given in Eq.~\ref{eq:matching coupling}. Since $\gamma_2<\gamma_1=10.4$, we adopt a smaller value $\gamma_2=7.0$ in the calculation and try to determine the parameters $\delta$ and $\zeta$ and discuss the phenomenological consequence. 

In Table~\ref{tab:para-new} all the parameters are listed. With the fixed parameters we can first calculate the partial decay widths for $K^*(1680)$ by treating it as the conventional $q\bar{q}$ states. We then introduce the hybrid mixing in the wavefunction to provide a possible solution.

\begin{table}[H]
    \centering
    \caption{ In our numerical studies the parameters of the QPC model are fixed, which include the $u/d$ and $s$ quark constituent mass $m_{q/s}$, the HO parameters $\beta$, QPC coupling strength $\gamma_1$ for the $q\bar{q}$ component, and $\gamma_2$ for the hybrid collinear mode transition, SU(3) breaking factor $R$, and the flavor singlet and octet mixing angle $\alpha_p$ for the $\eta$-$\eta'$ mixing. The mixing angle $\zeta$ between the conventional $q\bar{q}$ and the hybrid component within $K^*(1680)$ is fitted with the transverse mode transition fixed with $\delta=0.8,\ 1.0,\ 1.2$, and $-0.8$, which are labelled by the superscripts $a,\ b,\ c$, and $d$,  respectively.}\label{tab:para-new}
    \scalebox{0.9}{
    \begin{tabular}{c|cccccccccccc}
    \hline\hline
         &$m_q$ &$m_s$ &$\beta_P$ &$\beta_V$ &$\gamma_1$ &$\gamma_2$ &$R$  &$\alpha_P$ &$\zeta^a$ &$\zeta^b$ &$\zeta^c$ &$\zeta^d$\\
         \hline
         Value &$300$ MeV &$500$ MeV  &$550$ MeV &$450$ MeV &$10.4$ &$7$ &$m_q/m_s$ &$42^{\circ}$ &$9.79^{\circ}\pm 1.23^{\circ}$ &$8.28^{\circ}\pm 0.93^{\circ}$ &$7.15^{\circ}\pm 0.76^{\circ}$ &$171.11^{\circ}\pm 1.18^{\circ}$\\
         \hline
         &fixed &fixed &fixed &fixed &fixed &fixed &fixed&fixed &fitted &fitted &fitted &fitted \\
    \hline\hline
    \end{tabular}}
\end{table}

\subsection{Interpretation in the QPC model}

In Table~\ref{tab:QPC coupling} we listed the extracted effective couplings in the QPC model by treating $K^*(1680)$ as a $q\bar{q}$ vector state, namely, either $|q\bar{q}(2^3S_1)\rangle$, $|q\bar{q}(3^3S_1)\rangle$, or $|q\bar{q}(1^3D_1)\rangle$. One can see that the radial excitation of $|q\bar{q}(3^3S_1)\rangle$ has rather small couplings to the $PP$ and $VP$ channels. We will see that it means that $K^*(1680)$ is unlikely to be $|q\bar{q}(3^3S_1)\rangle$ state. 

To be more specific, the corresponding partial decay widths are listed in Table~\ref{tab:my_label} which can be compared with the experimental data. The calculated partial widths with $K^*(1680)$ as the pure $|q\bar{q}(2^3S_1)\rangle$ and $|q\bar{q}(3^3S_1)\rangle$ states indicate significant discrepancies with the experimental data. For $|q\bar{q}(2^3S_1)\rangle\to K^*\pi$ and $K\rho$, the calculated values can be regarded as in agreement with the experimental data within the uncertainties. But the partial widths for $K^*(1680)\to K\pi$ and $K\eta$ apparently cannot be explained by the transitions of 
$|q\bar{q}(2^3S_1)\rangle\to K\pi$ and $K\eta$, respectively. With the assignment of a pure $|q\bar{q}(3^3S_1)\rangle$ state, one sees that almost none of the decay channels can be accounted for.
In contrast, with the assignment of a pure $|q\bar{q}(1^3D_1)\rangle$ state a better agreement with the experimental data can be achieved in the $K\pi$, $K^*\pi$ and $K\rho$ channel. The results with $\gamma_1= 10.4$ adopted for the QPC coupling are also consistent with those reported in Ref.~\cite{Pang:2017dlw}. However, the calculated partial width for the $K\eta$ channel is one order of magnitude larger than the data. Also, the partial widths of the $K^*\pi$ and $K\rho$ channel are close to the lower bounds of the uncertainty ranges.

These results show that $K^*(1680)$ can hardly be interpreted as either the radial excitation states or the $D$-wave vector. One may wonder whether the $S-D$ mixing can provide a solution here. This is very much unlikely as shown by the pure state calculations. For the $|q\bar{q}(2^3S_1)\rangle-|q\bar{q}(1^3D_1)\rangle$ mixing, if the $S$-wave is dominant, it will require a constructive interference from the $D$-wave in the $PP$ channel. This will lead to an even larger partial width for the $K\eta$ channel, and be in contradiction with the data. 
If the $D$-wave is dominant, it will require a destructive interference from the $S$-wave to satisfy the $K\pi$ channel. However, a smaller and destructive $S$-wave will still result in a large partial decay width for the $K\eta$ channel. 

For the  $|q\bar{q}(3^3S_1)\rangle-|q\bar{q}(1^3D_1)\rangle$ mixing, the much smaller $|q\bar{q}(3^3S_1)\rangle$ coupling to the $PP$ and $VP$ channel means that it will be impossible to explain the small partial width for the $K\eta$ decay if the other channels can be explained. In short, one can conclude that the pure $q\bar{q}$ scenario is not sufficient for interpreting the experimental data though the data still bare large uncertainties. This makes the it a natural option to introduce the $q\bar{q}$ and hybrid mixing as a possible solution.

\begin{table}[H]
    \centering
    \caption{The coupling constants extracted in the QPC model for $K^*(1680)\to PP$ and $VP$, where $K^*(1680)$ is assumed to be the pure $|q\bar{q}(2^3S_1) \rangle$, $|q\bar{q}(3^3S_1)\rangle$, and $|q\bar{q}(1^3D_1) \rangle$ states, respectively.  Given that $K^*(1680)$ be the lowest energy hybrid, the $q\bar{q}$ component should be in the ground state $1^3S_1$. The corresponding coupling constants extracted in the QPC model at the mass of $K^*(1680)$ are listed in the fifth and sixth rows, where the subscripts ``1" and ``2" denote the different values $\gamma_1 = 10.4$ and $\gamma_2 = 7$ adopted in the calculation, respectively.}
    \begin{tabular}{c|cccccccc}
        \hline\hline
            &$K^+\pi^0$ &$K^+\eta$  &$K^+\eta^{\prime}$ &$K^{*+}\pi^0$ &$K^+\rho^0$ &$K^+\omega$ &$K^+\phi$ &$K^{*+}\eta$  \\
         \hline
         $g^{\text{QPC}}(2^3S_1)$ &$1.18$ &$2.16$ &$0.14$&$4.62$ &$5.32$ &$5.35$ &$5.64$ &$0.93$ \\
         \hline
         $g^{\text{QPC}}(3^3S_1)$ & $0.02$  & $-0.23$ & $-0.01$ & $-1.16$ & $-1.54$ & $-1.57$ & $-2.06$ & $-0.27$ \\
         \hline
         $g^{\text{QPC}}(1^3D_1)$ & $2.39$ & $3.52$ & $0.57$ & $-3.02$ & $-3.25$ & $-3.27$ & $-2.79$ & $-1.12$ \\
         \hline
         $g^{\text{QPC}}(1^3S_1)_1$ & $-4.66$ & $-7.12$ & $0.29$ & $-10.23$ & $-11.26$ & $-11.31$ & $-0.68$ & $-1.12$ \\
         \hline
         $g^{\text{QPC}}(1^3S_1)_2$ & $-3.14$ & $-4.79$ & $0.19$ & $-6.88$ & $-7.58$ & $-7.59$ & $-7.61$ & $-0.46$ \\
         \hline\hline
    \end{tabular}
    \label{tab:QPC coupling}
\end{table}

\begin{table}[H]
    \centering
    \caption{The partial decay widths (in MeV) based on different scenarios for $K^*(1680)$. The experimental results are listed in the second row. The values in the third, fourth, and fifth rows are obtained by treating $K^*(1680)$ as a pure $q\bar{q}$ state. The values in the sixth, seventh, eighth, and ninth rows are the results of considering the mixing between the $1^3D_1$ state and the hybrid state $|q\bar{q}\tilde{g}\rangle$. The superscripts $a,\ b,\ c$, and $d$ on $|q\bar{q}\tilde{g}\rangle$ correspond to the best fitting results for $\zeta$ when $\delta$ takes the different values of $0.8,\ 1.0,\ 1.2$, and $-0.8$, respectively. The last column shows the $\chi^2$ values for the four fitting schemes.}
    \scalebox{0.85}{
    \begin{tabular}{c|cccccccccc}
        \hline\hline
        Modes &$K\pi$ &$K\eta$ &$K\eta^\prime$ &$K^*\pi$ &$K\rho$ &$K\omega$ &$K\phi$ &$K^*\eta$&Tot &$\chi^2$\\
        \hline
        Expt. &$123.84^{+53.32}_{-47.82}$ &$4.48^{+5.84}_{-3.22}$ &- &$95.68^{+42.35}_{-43.39}$  &$100.48^{+56.04}_{-38.95}$ &- &seen &- &$320\pm 110$&-\\
        \hline
        pure $|q\bar{q}(2^3S_1)\rangle$ &35.95&26.76&0.03&135.77&142.02&46.90&16.62&0.76&404.82&-\\
        \hline
        pure $|q\bar{q}(3^3S_1)\rangle$ &0.008&0.31&$\sim0$&8.62&12.00&4.02&2.22&0.06&27.24&-\\
        \hline
    
        pure $|q\bar{q}(1^3D_1)\rangle$ &$147.18$ &$70.97$ &$0.50$ &$57.95$ &$53.21$ &$17.53$ &$4.06$ &$1.09$ &$352.48$ &-\\
        \hline
        $|q\bar{q}(1^3D_1)\rangle+|q\bar{q}\tilde{g}\rangle^a$ &$85.52\pm 7.07$ &$3.28\pm 4.68$ &$0.62\pm 0.02$ &$109.35\pm 7.13$ &$101.55\pm 6.73$ &$70.88\pm 8.76$ &$13.46\pm 1.49$ &$1.98\pm 0.13$ &$386.63\pm 21.85$ &1.12\\
        \hline
        $|q\bar{q}(1^3D_1)\rangle+|q\bar{q}\tilde{g}\rangle^b$ &$94.31\pm 5.48$ &$3.84\pm 4.54$ &$0.62\pm 0.02$ &$100.74\pm 5.21$ &$83.43\pm 4.91$ &$69.56\pm 7.59$ &$12.80\pm 1.23$ &$2.02\pm 0.12$ &$377.31\pm 18.11$ &0.67\\
        \hline
        $|q\bar{q}(1^3D_1)\rangle+|q\bar{q}\tilde{g}\rangle^c$ &$101.14\pm 4.57$ &$4.10\pm 4.46$ &$0.62\pm 0.02$ &$94.39\pm 4.15$ &$87.45\pm 3.91$ &$68.26\pm 6.99$ &$12.28\pm 1.08$ &$2.03\pm 0.11$ &$370.27\pm 16.13$ &0.49\\
        \hline
        $|q\bar{q}(1^3D_1)\rangle+|q\bar{q}\tilde{g}\rangle^d$ &$208.67\pm 8.25$ &$7.17\pm 4.13$ &$0.50\pm 0.01$ &$23.39\pm 3.66$ &$21.01\pm 3.39$ &$25.36\pm 1.06$ &$3.31\pm 0.11$ &$1.55\pm 0.07$ &$290.95\pm 1.91$ &8.88\\
        \hline\hline
    \end{tabular}}
    \label{tab:my_label}
\end{table}

\subsection{Interpretation with the $q\bar{q}$-hybrid mixing}

As discussed earlier, the hybrid decays into $PP$ and $VP$ via Fig.~\ref{Fig:FT} (a) can be described by the QPC model since the quark pair created by the gluelump will involve multi-soft-gluon exchanges which in the end will lead to a similar behavior as the $^3P_0$ decay. It makes the process of Fig.~\ref{Fig:FT} (a) indistinguishable from an initial $|1^3S_1(q\bar{q})\rangle$. Note that  the effective coupling $g_1$ is extracted by matching it to the QPC model, and assuming that it is connected to the ground state $|1^3S_1(q\bar{q})\rangle$ coupling to $PP$ and $VV$ but with the mass at the $K^*(1680)$. However, since the physical mass, i.e. the mass of $K^*(1680)$, is far away from the ground state vector meson, it suggests that some suppression effects, such as the off-shell form factors, should be present. In this sense the coupling $\gamma_2$ defined in the QPC model (in connection with the effective coupling $g_1$) should not be larger than the ground state $q\bar{q}$ coupling $\gamma_1$ adopted for the QPC model. Furthermore, the hybrid coupling in the collinear mode is correlated with the $q\bar{q}$-hybrid mixing angle through the wavefunction. Due to lack of further constraints from either experiment or theory we have to survey the correlation among these three parameters, i.e. $\gamma_2$, $\delta$ and $\zeta$.

To proceed, we assume that the $q\bar{q}$ component in the $K^*(1680)$ wavefunction is $|q\bar{q}(1^3D_1)\rangle$, which will mix with the hybrid component. With the the $K^*(1680)$ wavefunction $|K^*(1680)\rangle=\cos\zeta |q\bar{q}(1^3D_1)\rangle+\sin\zeta |q\bar{q}\tilde{g}\rangle$, the transition amplitude will be described by Eq.~(\ref{tot-trans-amp}), where the transition of the $|q\bar{q}(1^3D_1)\rangle$ component is described by the QPC mode, while the hybrid component decay will be parametrized out by the FT model as formulated earlier. It should be noted again that the collinear mode can be connected to the ground state $q\bar{q}$ decay in the QPC model instead of the excited ones. Therefore, when implement the QPC model to describe the transition of $|q\bar{q}\tilde{g}\rangle\to PP$ and $VP$ via the collinear mode, its analogue to the QPC model is via the $|q\bar{q}(1^3S_1)\rangle$ configuration which requires  $\gamma_2<\gamma_1$.

\begin{figure}[H]
    \centering
    \includegraphics[width=0.9\linewidth]{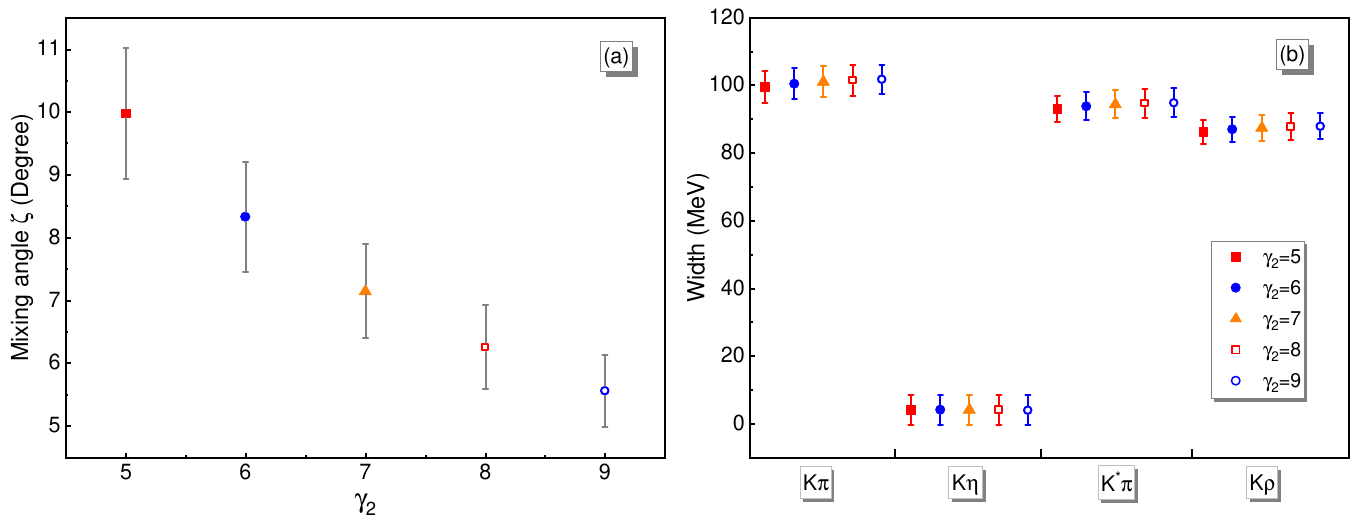}
    \caption{(a) The fitted mixing angle $\zeta$ for different $\gamma_2$ values: $\gamma_2=5,\ 6,\ 7,\ 8,$ and $9$ with $\delta=1.2$. (b) Partial decay widths for the four decay modes $K\pi$, $K\eta$, $K^*\pi$, and $K\rho$ are fitted by the mixing angle $\zeta$ with $\gamma_2=5,\ 6,\ 7,\ 8$, and $9$ and $\delta=1.2$.}
    \label{fig:correlation}
\end{figure}

With $\gamma_1= 10.4$ adopted for the QPC coupling the partial decay widths of the pure $|q\bar{q}(1^3D_1)\rangle$ state have been listed in Table~\ref{tab:my_label}. 
Actually, since the term from Fig.~\ref{Fig:FT} (b) only contributes to those channels containing isoscalar states, e.g. $K\eta$, $K^*\eta$, etc., the correlation between $\gamma_2$ and $\zeta$ can be investigated in decay channels which only involve isospin-none-zero channels, i.e. $K\pi$, $K^*\pi$, and $K\rho$. In principle, one can first determine $\gamma_2 \ (<\gamma_1)$ and $\zeta$ by fitting the data for the $K\pi$, $K^*\pi$, and $K\rho$ decay channels. Then, with the fixed $\gamma_2$ and $\zeta$, the coupling ratio $\delta$ can be determined by the $K\eta$ channel. However, since there are still large uncertainties with the experimental data, and the partial width of $K^*(1680)\to K\eta$ is much smaller than other channels, the effects arising from the hybrid transverse mode do not contribute significantly to the $\chi^2$ in some ranges of the $\delta$ values.
Therefore, instead of determining $\gamma_2$ and $\zeta$ first, we first restrict the range of $\delta$ by arguing that $|\delta|\simeq 1$. Our numerical study shows that the constraints of $\gamma_2 <\gamma_1$ and $|\delta|\simeq 1$ help to find reasonable fitting results. It shows that the best parameter space can be obtained with $\delta\simeq 1.2$. In Fig.~\ref{fig:correlation}, we show the correlation between $\gamma_2$ and $\zeta$ by fixing $\delta=1.2$ and for different values of $\gamma_2=5, \ 6, \ 7, \ 8, \ 9$, the mixing angle $\zeta$ is determined by fitting the four measured channels. In Fig.~\ref{fig:correlation} (a) it shows that with the increase of $\gamma_2$, the mixing angle $\zeta$ drops which actually shows that correlation between $\gamma_2$ and $\zeta$ via the hybrid amplitude proportional for $\gamma_2\sin\zeta$. With $\gamma_2$ fixed with larger value it will allow a better determination of $\zeta$. Thus, the smaller errors for the fitted $\zeta$ with larger $\gamma_2$ are understandable. 

In Fig.~\ref{fig:correlation} (b) we present the partial decay widths for different channels with the parameter values shown in Fig.~\ref{fig:correlation} (a). The results do not show significant differences, which indicates again the correlation between $\gamma_2$ and $\zeta$ with the fixed value of $\delta=1.2$. Although the $\delta$ term only contribute to those channels involving the isoscalar pseudoscalar final state, we can also see its impact on the numerical fitting. In  Table~\ref{tab:my_label} we list the  best fitting results for $\zeta$ when $\delta$ takes the different values of $0.8,\ 1.0,\ 1.2$, and $-0.8$, which are labelled by the superscripts $a,\ b,\ c$, and $d$, respectively. The fitting quality is shown by the $\chi^2$ value in the last column.

As shown in Table~\ref{tab:my_label}, the best fitting is the case with $\gamma_2=7$ and $\delta=1.2$. It shows that the hybrid mixing will bring down the partial width of the $PP$ channels for the pure $|q\bar{q}(1^3D_1)\rangle$ decay, and enhance those $VP$ channels. In particular, the $K\eta$ channel receives the destructive interference and the partial width drops to be within the experimental range. It is interesting to note that the transition amplitudes of $|q\bar{q}(1^3D_1)\rangle\to PP$ and $VP$ have different signs arising from the spin factors. However, for the ground state $q\bar{q}(1^3S_1)$ these two transitions have the same sign. This feature turns to be a crucial feature to account for the pattern for the $K^*(1680)\to PP$ and $VP$ decays. Note that, only the $S-D$ mixing cannot give a good fitting result as that from the scheme ``$c$" in Table~\ref{tab:my_label}.

From Fig.~\ref{fig:correlation} (a) the fitted mixing angle  $\zeta=7.15\degree\pm0.76\degree$, which turns out to be a rather small value. It suggests that the behavior of $K^*(1680)$ is still dominated by the conventional $q\bar{q}$ configuration, while the hybrid component is quite small. However, as shown by our numerical survey, such a small hybrid component is crucial for describing the $K^*(1680)$ decay pattern.

\subsection{Further phenomenological implication of the $q\bar{q}$-hybrid mixing}

The extracted small mixing angle $\zeta=7.15\degree\pm0.76\degree$ allows us to understand more about the $q\bar{q}$-hybrid mixing in the vector strange sector. On a more general basis we can express the eigen equation of two configuration mixing as
\begin{eqnarray}
    \left(\begin{array}{cc}
        \hat{H}_q & \hat{H}_\Delta \\
        \hat{H}_\Delta & \hat{H}_h 
    \end{array}\right)
    \left(\begin{array}{c}
        \alpha \psi_q \\ \beta\psi_h 
        \end{array}\right) &=& E\left(
        \begin{array}{c}
        \alpha \psi_q \\ \beta\psi_h 
        \end{array}\right) \ ,
\end{eqnarray}
where $\hat{H}_q$ and $\hat{H}_h$ are the Hamiltonians for the pure $q\bar{q}$ ($\psi_q$) and pure hybrid ($\psi_h$) systems, respectively; $\hat{H}_\Delta$ is the transition operator for the $\psi_q$ and $\psi_h$ transition; $\alpha$ and $\beta$ are probability amplitudes for the physical state to be in the pure states of $\psi_q$ and $\psi_h$, respectively. The normalisation condition requires $\alpha^2+\beta^2=1$. 

Given the eigen values, $E_q\equiv \langle \psi_q|\hat{H}_q|\psi_q\rangle$ and $E_h\equiv \langle \psi_h|\hat{H}_h|\psi_h\rangle$, and with the transition amplitude $\Delta\equiv \langle \psi_q|\hat{H}_\Delta|\psi_h\rangle=\langle \psi_h|\hat{H}_\Delta|\psi_q\rangle$, the eigen equation can be written as 
\begin{eqnarray}
    \left(\begin{array}{cc}
        E_q & \Delta \\
        \Delta & E_h 
    \end{array}\right)
    \left(\begin{array}{c}
        \alpha  \\ \beta
        \end{array}\right) &=& E\left(
        \begin{array}{c}
        \alpha  \\ \beta 
        \end{array}\right) \ ,
\end{eqnarray}
of which the eigen-value is given by the vanishing determinant of coefficient
\begin{eqnarray}
    E&=&\frac 12\left[(E_q+E_h)\pm\sqrt{(E_q+E_h)^2-4(E_q E_h-\Delta^2)}\right] \ ,
\end{eqnarray}
where we define $E_L$ and $E_H$ corresponding to the lower (``$-$") and higher (``$+$") solution, respectively. Substituting these two solutions back into the eigen equation, we can obtain the physical wave function $|\Psi\rangle=\alpha |\psi_q\rangle+\beta|\psi_h\rangle$.

In our convention the extracted physical state is expressed as $|K^*(1680)\rangle=\cos\zeta |q\bar{q}(1^3D_1)\rangle+\sin\zeta |q\bar{q}\tilde{g}\rangle$, where $\psi_q=|q\bar{q}(1^3D_1)\rangle$ and $\psi_h=|q\bar{q}\tilde{g}\rangle$, $\alpha=\cos\zeta$ and $\beta=\sin\zeta$ can be established. With $\zeta=7.15\degree\pm0.76\degree$ it is easy to find out that $K^*(1680)$ as the higher mass state $E_H$ can satisfy the requirement. The explicit expression for the mixing angle can be obtained
\begin{eqnarray}
    \alpha &=&\cos\zeta =\frac{\Delta}{E_H-E_q}\sqrt{\frac{E_H-E_q}{2E_H-(E_q+E_h)}} \ ,\\
    \beta &=&\sin\zeta = \sqrt{\frac{E_H-E_q}{2E_H-(E_q+E_h)}} \ ,
\end{eqnarray}
where both $\alpha$ and $\beta$ are positive, and $E_L<(E_q+E_h)/2<E_H$ is apparent. 

One notices that to satisfy the small mixing angle $\zeta=7.15\degree\pm0.76\degree$, the requirement of $\alpha>>\beta$ leads to $\Delta/(E_H-E_q)>>1$. It suggests that the physical state $K^*(1680)$ has the mass close to the pure $|q\bar{q}(1^3D_1)\rangle$ state. Meanwhile, the mass difference between the physical state and the pure $|q\bar{q}(1^3D_1)\rangle$ is much smaller than the transition element between the pure $|q\bar{q}(1^3D_1)\rangle$ and pure hybrid state $|q\bar{q}\tilde{g}\rangle$. 

Following this scenario, the lower physical state in this mixing scheme can be obtained: $\Psi_L=-\beta\psi_q+\alpha\psi_h$, and the eigen value is $E_L=\frac 12 [(E_q+E_h)-\sqrt{(E_q-E_h)^2+4\Delta^2}]$. Recall that $\Delta/(E_H-E_q)>>1$, i.e. $\Delta >>(E_H-E_q)=(E_h-E_L)$, and the pure $|q\bar{q}(1^3D_1)\rangle$ and pure hybrid state $|q\bar{q}\tilde{g}\rangle$ are possibly close to each other~\cite{Qiu:2022ktc}. We approximate $E_L\simeq \frac 12 [(E_H+E_L)-2\Delta]$, i.e. $E_L\simeq E_H-2\Delta$. It means that the lower physical state will be pulled down to be much lower than the mass of the pure hybrid. Also note that in the lower mass region we have observed $K^*(1410)$ which is very different from the quark model expectation. If this corresponds to the physical state which is dominated by the hybrid component~\cite{Barnes:2002mu}, we estimate that $\Delta\simeq (E_H-E_L)/2\simeq (1680-1410)/2=135$ MeV, which is a typical scale for strong transitions in the non-perturbative regime. This makes it interesting to further study the strange vector spectrum including  $K^*(1410)$, $K^*(1680)$ and other vector states. It also transfers an important message that the strange partner of the hybrid nonet may be affected strongly by the mixing mechanism. A combined understanding of the hybrid states and the excited vector states in the strange sector should be pursued.

\section{Summary}\label{Sec:IV}

In this work we have carried out  a study of the $K^*(1680)$ state via  its strong decays into two-body final states within the FT model and QPC model. We find that its decay pattern cannot be described by the $q\bar{q}$ scenario based on the QPC model calculations. We show  qualitatively that the $S-D$ mixing cannot interpret the decay pattern of the $K^*(1680)$ decays into $PP$ and $VP$ channels. Taking into account that the strange vector states may mix with the hybrid states with $J^{P(C)}=1^{-(+)}$, we consider $K^*(1680)$ to be a physical state of the $q\bar{q}$-hybrid mixing, and a reasonable description of the two-body decay pattern can be obtained. Although the experimental data still have large uncertainties, the numerical study suggests a crucial role played by the $q\bar{q}$-hybrid mixing mechanism. It is interesting to find out that $K^*(1680)$ has a dominant $q\bar{q}$ component, but needs a relatively small hybrid component in the wave function. We have also discussed the phenomenological consequence of such a scenario. It may imply that $K^*(1410)$ contains a relatively large hybrid component. This may explain that its mass cannot be accommodated by the conventional quark model multiplets. Further studies of the vector spectrum of the light strange mesons are needed. Our study can provide a guidance for future search for the strange hybrid at BESIII, LHCb and Belle-II experiments.

\begin{acknowledgments}
Useful discussions with Profs. Xiao-Yan Shen and Bei-Jiang Liu concerning the BESIII experimental results are acknowledged. This work is supported, in part, by the National Natural Science Foundation of China under Grant Nos. 12235018, 12265010, and 12505098, and the Project of Guizhou Provincial Department of Science and Technology under Grant Nos. MS[2025]219, CXTD[2025]030, ZD[2026]168, and QN[2025]178, and the project of Guizhou Normal University under Grant No. GZNUD[2025]05. Also, Samee Ullah is thankful to Chinese Scholarship Council (CSC) for the award of PhD scholarship to complete this work.
\end{acknowledgments}

\end{document}